\documentclass[aps,prd,floatfix]{revtex4} 

\usepackage{graphics}
\usepackage{graphicx}

\usepackage{latexsym}
\usepackage[cp866]{inputenc}
\usepackage[english]{babel}

\input epsf

\begin{document}

\title{\boldmath Electronic width of the $\psi(3770)$ resonance interfering with the background}
\author{N.N. Achasov \footnote{achasov@math.nsc.ru}
and G.N. Shestakov \footnote{shestako@math.nsc.ru}
}\affiliation{Laboratory of  Theoretical Physics, S.L. Sobolev
Institute for Mathematics, 630090, Novosibirsk, Russia}


\begin{abstract}

Methods for extracting the $\psi(3770)\to e^+e^-$ decay width from
the data on the reaction cross section $e^+e^-\to D\bar D$ are
discussed. Attention is drawn to the absence of the generally
accepted method for determining $\Gamma_{\psi(3770)e^+e^-}$ in the
presence of interference between the contributions of the
$\psi(3770)$ resonance and background. It is shown that the model
for the experimentally measured $D$ meson form factor, which
satisfies the requirement of the Watson theorem and takes into
account the contribution of the complex of the mixed $\psi(3770)$
and $\psi(2S)$ resonances, allows us to uniquely determine the value
of $\Gamma_{\psi(3770)e^+e^-}$ by fitting. The
$\Gamma_{\psi(3770)e^+ e^-}$ values found from the data processing
are compared with the estimates in the potential models.

\end{abstract}

\maketitle

\section{INTRODUCTION}

The charmonium state $\psi(3770)$ \cite{PDG20} predicted in the
mid-seventies is considered as the $1^3D_1$ state of the $c\bar c$
system with small admixtures of $n^3S_1$ states [mainly $\psi(2S)$]
\cite{Ei75,Ei76,Ei04,Ei80,Ei06,Ei08,No78,Ro01,Ro05, Ja77,HTO84}. In
$e^+e^-$ collisions, the $\psi(3770)$ resonance is observed in the
form of the resonant enhancement, with a width of about 30 MeV,
located between the $D\bar D$ ($2m_D\approx3.739$ GeV) and $D\bar
D^*$ ($m_D+m_{D^*} \approx 3.872$ GeV) production thresholds. The
sizeable width of the $\psi(3770)$ resonance is due to its strong
decays into $D\bar D$ meson pairs. Indeed, the fraction of the
radiative decays $\psi(3770)\to\gamma \chi_{ cJ=0,1,2}$, $\gamma
\eta_c$, $\gamma\eta_c(2S)$ is less than 1.5\%, and the fraction of
the $\psi(3770)\to J/\psi \pi^+ \pi^-$, $J/ \psi\pi^0 \pi^0$ and
$J/\psi \eta$ decays is less than $0.5\%$ \cite{PDG20}. The total
width of the Zweig forbidden decays $\psi(3770)\to light\ hadrons$
must be comparable from the theoretical point of view with the
corresponding decay widths of the $J/\psi$ and $\psi(2S)$ resonances
located under the $D\bar D$ threshold. In order of magnitude, it can
be about 100 keV, which is less than $0.5\%$ of the total decay
width of the $\psi(3770)$ meson. For almost ninety decay channels,
$\psi(3770)\to light\ hadrons$ are known only upper limits (some of
which are rather high) \cite{PDG20}. Only the branching ratio of the
decay $\psi(3770)\to \phi\eta$ is definitely known,
$\mathcal{B}(\psi(3770)\to\phi\eta) =(3.1\pm0.7)\times 10^{-4}$
\cite{PDG20}.

The charmonium state $\psi(3770)$ was investigated in $e^+e^-$
collisions by the MARK-I \cite{Ra,Pe}, DELCO \cite{Ba}, MARK-II
\cite{Sch}, BES \cite{Ab1,Ab2,Ab3,Ab3a,Ab4,Ab5, Ab6,Ab7,Ab8}, CLEO
\cite{Be1,Do, Be2}, {\it BABAR} \cite{Au1,Au2}, Belle \cite{Pa}, and
KEDR \cite{An2} Collaborations. The $\psi(3770)$ production was also
observed in the $B^+\to D\bar DK^+$ decays by the Belle \cite{Ch04,
Br08}, {\it BABAR} \cite{Au08,Le15}, and LHCb \cite{Aa20}
Collaborations. Full compilation of the $\psi(3770)$ production
experiments is contained in the review of the Particle Data Group
(PDG) \cite{PDG20}. The unusual shape of the $\psi(3770)$ resonance
peak, discovered in many experiments \cite{Ab3a,Ab4,Ab6,Ab7,Ab8,
Au1,Au2,Pa,An2}, naturally became the subject of many-sided
theoretical analyses (see, for example, Refs. \cite{Ya,LQY,ZZ,AS12,
AS13,CZ,Li,CL14a, CL14b,DMW,ST,CG}). The following circumstance is
also of additional interest. According to the CLEO data \cite{Be1,
Do,Be2}, the value of the non-$D\bar D$ component in the decay width
of the $\psi(3770)$ resonance is negligible. At the same time, the
BES analysis \cite{Ab2,Ab3,Ab4,Ab5} does not exclude a noticeable
non-$D\bar D$ component. According to the theoretical estimates
\cite{He08,He10}, the non-$D\bar D$ decay branching ratio of
$\psi(3770)$ could reach about 5\%. The authors of Refs.
\cite{He08,He10} note that this result does not contain evidence in
favor of BES or CLEO results, and urge doing more precise
measurements on both inclusive and exclusive non-$D\bar D$ decays of
$\psi(3770)$ in the future. Unfortunately, this contradiction has
not yet been resolved. As a result, the PDG \cite{PDG20} gives the
following value for the $D\bar D$ component: $\mathcal{B}(\psi(3770)
\to D\bar D)=[\mathcal{B}(\psi(3770)\to D^+D^-)=(52\pm^4_5)\%]+[
\mathcal{B}(\psi (3770)\to D^0\bar D^0)=(41\pm4)\%]=(93\pm^8 _9)\%$.
Theoretical considerations combined with the CLEO data \cite{Be1,Do,
Be2} suggest that the dominance of the $\psi(3770)\to D\bar D$ decay
can be at the level of $97\%-98\%$. In what follows, we will
consider $\psi(3770)$ to be an almost elastic resonance coupled to
the $D\bar D$ decay channels and apply this assumption to describe
its line shape and determine its electronic decay width,
$\Gamma_{\psi(3770)e^+e^-}$.

This paper is organized as follows. Section II gives a brief
overview of the commonly used methods for describing the
$\psi(3770)$ resonance and the definitions of $\Gamma_{\psi(3770)
e^+ e^-}$, particularly those selected by the PDG \cite{PDG20} for
calculations fitting $(0.262\pm0.018)$ keV and average $(0.256
\pm0.016)$ keV values of $\Gamma_{\psi(3770)e^+e^-}$. Attention is
drawn to the fact that some seemingly natural parametrizations of
the cross section $\sigma( e^+e^-\to D\bar D)$, taking into account
the interference of the $\psi(3770)$ resonance and background, do
not allow us to determine the value of $\Gamma_{\psi(3770)e^+ e^-}$
uniquely. In Sec. III, we apply to the description of the reaction
cross section $\sigma( e^+e^-\to D\bar D)$ the model for the
isoscalar form factor of the $D$ meson, which takes into account the
contributions of the $\psi(3770)$ and $\psi(2S)$ resonances mixed
due to their coupling with the $D\bar D$ decay channels. The model
satisfies the requirement of the unitarity condition or the Watson
theorem \cite{Wa52} and allows us to unambiguously determine the
value of $\Gamma_{\psi(3770)e^+e^-}$ from the data by fitting. Our
analysis substantially develops the approach proposed in Refs.
\cite{AS12,AS13} by consistently taking into account the finite
width corrections in the resonance propagators and clarifying their
important role. In Sec. IV, we compare the values of
$\Gamma_{\psi(3770)e^+e^-}$ found from phenomenological data
processing with theoretical estimates in potential models and
briefly state our conclusions.


\section{\boldmath Parametrizations of the $\psi(3770)$ resonance structure}

In many experimental works, the cross section of the reaction
$e^+e^-\to D\bar D$ in the $\psi(3770)$ resonance region was
described with minor modification by the following formula \cite{Ra,
Pe,Ba,Sch,Ab1,Ab2,Ab3, Ab3a,Ab4,Ab5,Be1} [for short, $\psi(3770)$ is
also denoted as $\psi''$ below]:
\begin{equation}\label{Eq1} \sigma_{\psi''}(e^+e^-\to D\bar
D;\,s)=\frac{12\pi\Gamma_{\psi''e^+ e^-}\Gamma_{\psi''D\bar D}(s)}{
(m^2_{\psi''}-s)^2+(m_{\psi'' }\Gamma^{tot}_{\psi''}(s))^2},
\end{equation} where $s$ is the invariant mass squared of the $D\bar D$
system, $m_{\psi''}$, $\Gamma_{\psi''e^+ e^-}$, $\Gamma_{\psi''D
\bar D}(s)$, and $\Gamma^{tot}_{\psi''}(s)$ are the mass,
electronic, $D\bar D$, and total decay widths of $\psi''$,
respectively. The energy-dependent width $\Gamma_{\psi''D\bar D}(s)$
[dominating in $\Gamma^{tot}_{\psi''}(s)$] was taken in the form
\begin{equation}\label{Eq2} \Gamma_{\psi''D\bar D}(s)=G^2_{\psi''}
\left(\frac{p^3_0(s)}{1+r^2p^2_0(s)}+\frac{p^3_+(s)}{1+r^2p^2_+
(s)}\right), \end{equation} where $p_0(s)$\,=\,$\sqrt{s/4-
m^2_{D^0}}$ and $p_+(s)$\,=\,$\sqrt{s/4-m^2_{D^+}}$ are the $D^0$
and $D^+$ momenta in the $\psi''$ rest frame, $r$ is the $D\bar D$
interaction radius \cite{BW52}, and $G_{\psi''}$ is the coupling
constant of the $\psi''$ with $D\bar D$.

For the solitary $\psi''$ resonance, there is no problem with
determining $\Gamma_{\psi''e^+ e^-}$ by fitting the data using Eqs.
(\ref {Eq1}) and (\ref {Eq2}). Discrepancy between the values found
by different Collaborations ($\Gamma_{\psi''e^+e^-}= 345\pm 85$ eV
\cite{Pe}, $180\pm60$ eV \cite{Ba}, $276\pm50$ eV \cite{Sch},
$279\pm11\pm13$ eV \cite{Ab4}, $220\pm50$ eV \cite{Ab6,PDG20},
$204\pm3^{+41}_{-27}$ eV \cite{Be1}) is mainly related to the
difference in the collected raw data and uncertainties in the cross
section normalization.

With increasing accuracy of measurements, there appeared to be
indications of an unusual (anomalous) shape of the $\psi(3770)$ peak
in the $e^+e^-\to\psi''\to hadrons$ and $e^+e^-\to\psi''\to D\bar D$
reaction cross sections, i.e., on possible interference effects that
occur directly in the $\psi(3770)$ resonance region \cite{Ab3a,Ab4,
Ab6,Ab7,Ab8, Au1,Au2,Pa,An2}. In particular, there is a deep dip in
the $D\bar D$ production cross section near $\sqrt{s}\approx3.81$
GeV \cite{Ab3a, Ab4,Au1,Au2,Pa} which strongly distorts the right
wing of the $\psi''$ resonance. Such a dip is difficult to describe
using Eqs. (\ref{Eq1}) and (\ref{Eq2}) for a solitary $\psi''$
resonance contribution. In Ref \cite{AS12}, we showed that the
description of the data \cite{Ab3a,Ab4,Au1,Au2,Pa,Be2} with the use
of these formulas turns out to be unsatisfactory for any values of
the parameter $r$. In addition, by performing the analytical
continuation of the amplitudes $e^+e^-\to \psi''\to D^0\bar D^0$ and
$e^+e^-\to\psi''\to D^+D^-$ corresponding to the parametrizations
(\ref{Eq1}) and (\ref{Eq2}) below the $D\bar D$ thresholds, it is
easy to make sure that they have spurious poles and left cuts due to
the $P$-wave Blatt and Weisskopf barrier penetration factors,
$1/[1+r^2p^2_{0,+}(s)]$ \cite{BW52}. For example, for
$r\approx1\,\mbox{fm}\approx5\, \mbox{GeV}^{-1}$, the indicated
singularities appear at about 20 MeV below the $D\bar D$ thresholds.
In the next section, we show that taking into account the finite
width corrections in the resonance propagators allows us to
eliminate these singularities.

If we are not dealing with a solitary resonance, but with a complex
of the mixed resonance and background contributions, then a
practical question arises about the way of describing it as a whole
and the possibilities of adequately determining the individual
characteristics of its components. In what follows, we will talk
about the process $e^+e^-\to D\bar D$, in which the isoscalar
electromagnetic form factor of the $D$ meson $F^0_D(s)$ is measured.
The sum of the $e^+ e^-\to D\bar D$ reaction cross sections is
expressed in the terms of $F^0_D(s)$ as follows:
\begin{equation}\label{Eq3} \sigma(e^+e^-\to D\bar D;\,s)=
\frac{8\pi\alpha^2}{3s^{5/2}}\left|F^0_D(s)\right|^2\left[p^3_0(s)+
p^3_+(s)\right]\,,\end{equation} where $\alpha$\,=\,$e^2/4\pi
$\,=\,1/137. Here we neglect the isovector part of the $D$ meson
form factor and do not touch on the question about the isospin
symmetry breaking. The KEDR Collaboration \cite{An2}, analyzing
their own data on the $e^+e^-\to D\bar D$ cross section, showed that
taking into account the interference between the $\psi(3770)$
resonance and background contributions affects the value$^,$s
resonance parameters and therefore the corresponding results cannot
be directly compared with those obtained ignoring this effect. In
addition, in Ref. \cite{An2}, within the framework of the accepted
parametrization for $F^0_D(s)$, two essentially different solutions
were obtained for the production amplitude of the $\psi(3770)$ state
and its phase relative to the background (see also \cite{ST}). These
two solutions lead to the same energy dependence of the $e^+e^-\to
D\bar D$ cross section and are indistinguishable by the $\chi^2$
criterion. Ambiguities of this type in the interfering resonance
parameter determination were found in Ref. \cite{Bu} (see also
\cite{Zh,Yu}). The PDG used one of the KEDR solutions \cite{An2}
[see Eq. (\ref{Eq8}) below] to determine the value of
$\Gamma_{\psi''e^+ e^-}=(0.262\pm0.018)$ keV \cite{PDG20}, together
with the above results from other works \cite{Ba,Sch,Ab4,Ab6,Be1}
(in which the interference was not taken into account).

Let us illustrate the ambiguity of the choice of the resonance
parameters with a simple example. Consider a model of the reaction
amplitude $e^+e^-\to h\bar h$ (where $h$ and $\bar h$ are hadrons),
which takes into account the resonance and background contributions;
\begin{equation}\label{Eq4}
F(E)=\frac{A_xe^{i\varphi_x}}{M-E-i\Gamma/2}+B_x
\end{equation}
Here, $E$ is the energy in the $h\bar h$ center-of-mass system, $M$
is the mass, $\Gamma$ is the energy-independent width of the
resonance, and $A_x$, $\varphi_x$, and $B_x$ are the real
parameters. At fixed $M$ and $\Gamma$, there are two solutions for
$A_x$, $\varphi_x$, and $B_x$ \cite{Bu}:
\begin{equation}\label{Eq5}\mbox{(I)}\quad A_x=A,\ \ B_x=B,\ \
\varphi_x=\varphi,\end{equation}
\begin{eqnarray}\label{Eq6}\mbox{(II)}\quad A_x=\sqrt{A^2-2AB\Gamma\sin
\varphi+B^2\Gamma^2},\ \ B_x=B,\ \  
\tan\varphi_x=-\tan\varphi+B\Gamma/(A\cos
\varphi),\qquad\end{eqnarray} which yield the same cross section as
a function of energy, $\sigma(E)$\,=\,$|F(E)|^2$, and different
amplitude, $A_x$, and phase, $\varphi_x$. For example, if
$M$\,=\,3.77 GeV, $\Gamma$\,=\,0.03 GeV $A$\,=\,0.045 nb$^{1/2}$GeV,
$\varphi $\,=\,0, and $B$\,=\,1.5 nb$^{1/2}$ for solution (I), then,
for solution (II), $A_x$\,=\,$\sqrt{2}A$ and $\varphi_x$\,=\,$
\pi/4$. Since $A_x\sim\sqrt{\Gamma_{e^+e^-}\Gamma}$, the values of
the electronic decay width of the resonance $\Gamma_{e^+ e^-}$
differ by a factor of two for solutions (I) and (II).

The similar form factor parametrization was used to determine the
$\psi(3770)$ resonance parameters in Ref \cite{An2}:
\begin{equation}\label{Eq7} F^0_D(s)=F^{\psi(3770)}(s)e^{i\phi}+
F^{\mbox{\scriptsize{N.R.}}}(s),\end{equation} where $F^{\psi(3770)}
(s)$ is the Breit-Wigner $P$-wave resonance amplitude,
$F^{\mbox{\scriptsize{N.R.}}}(s)$ is the background amplitude, and
$\phi$ is their relative phase. $F^{\mbox{\scriptsize{N.R.}}}(s)=
F^{\psi(2S)}(s)+F_0$ takes into account the contribution of the
right wing of the nearest resonance $\psi(2S)$ with the mass of
3.686 GeV and the additional constant contribution $F_0$. Two
solutions indistinguishable in $\chi^2$ are \cite{An2}:
\begin{equation}\label{Eq8}\mbox{(I)}\quad\Gamma_{\psi''e^+
e^-}=160^{+78}_{-58}\,\ \mbox{eV},\quad
\phi=(170.7\pm16.7)^\circ,\end{equation}
\begin{eqnarray}\label{Eq9}\mbox{(II)}\quad\Gamma_{\psi''e^+
e^-}=420^{+72}_{-80}\,\ \mbox{eV},\quad
\phi=(239.6\pm8.6)^\circ.\end{eqnarray}

Thus, parametrizations of types (\ref{Eq4}) and (\ref{Eq7}),
preserving at first glance the usual way of determining the
individual characteristics of the $\psi''$ resonance (for example,
its electronic width), do not allow to do this unambiguously by
fitting. If one of the values of $\Gamma_{\psi''e^+ e^-}$ from Eqs.
(\ref{Eq8}) and (\ref{Eq9}) agrees with some theoretical estimate of
$\Gamma_{\psi''e^+e^-}$, then it does not yet mean the  validity of
Eq. (\ref{Eq7}), which contains the phase $\phi$ of unknown origin
and does not take into account the transition amplitude between the
background and resonance through the common $D\bar D$ intermediate
states.

However, just in the case of the $\psi''$ resonance, the above
difficulties can be avoided if we take into account the requirement
of the unitarity condition. As noted above, the $\psi''$ is the
elastic resonance in a good approximation. But in the elastic region
(between $D\bar D$ and $D\bar D^*$ thresholds) with a width of about
141 MeV, the unitarity condition requires that the phase of the form
factor $F^0_D(s)$ coincide with the phase $\delta^0_1(s)$ of the
strong $P$-wave $D\bar D$ elastic scattering amplitude
$T^0_1(s)=e^{\delta^0_1(s)}\sin \delta^0_1(s)$ in the channel with
isospin $I=0$, i.e.,
\begin{equation}\label{Eq10} F^0_D(s)=e^{i\delta^0_1(s)}\mathcal{F}^0_D
(s)\,,\end{equation} where $\mathcal{F}^0_D(s)$ and $\delta^0_1(s)$
are the real functions of energy \cite{Wa52}. It is clear that
formulas (\ref{Eq4}) and (\ref{Eq7}) contradict the unitarity
requirement since the phase of the form factor determined by them
depends on the ratio of the background and resonance coupling
constants with $e^+e^-$, on which $\delta^0_1(s)$ is obviously
independent.

In the next section, we apply to the description of the data on the
reaction $e^+e^-\to D\bar D$ a simple model of the form factor
$F^0_D(s)$, which satisfies the requirement of the unitarity
condition for the case of the mixed $\psi''$ and $\psi(2S)$
resonances and allows by fitting to uniquely determine the value of
$\Gamma_{ \psi''e^+e^-}$. Our analysis is an advancement of that
which is suggested earlier in \cite{AS12,AS13}.


\section{\boldmath The $D$ meson electromagnetic form factor in
the $\psi(3770)$ region}
\subsection{{\boldmath The solitary $\psi''$ resonance}}

Consider a model that takes into account in the form factor $F^0_D
(s)$, amplitude $T^0_1(s)$, and the contributions of two resonances,
$\psi''$ and $\psi(2S)$, that are close to each other, strongly
coupled only to $D\bar D$ decay channels, and are mixing with each
other due to transitions $\psi''\to D\bar D\to\psi(2S)$ and vice
versa. However, we first write down the contribution of $\psi''$ to
$F^0_D(s)$ in the spirit of the vector dominance model
\cite{FN1,GS,RP,BM}, ignoring its mixing with $\psi(2S)$;
\begin{equation}\label{Eq11} F^0_D(s)=
F^{\psi''}_D(s)=\frac{C_{\psi''}}{\widetilde{D}_{\psi''}(s)}=
\frac{C_{\psi''}}{m^2_{ \psi''}-s-h_{\psi''}(s)-i\sqrt{s} \Gamma_{
\psi''D\bar D}(s)}, \end{equation} where $C_{\psi''}$ is an
$s$-independent constant, $\widetilde{D}_{\psi''}(s)$ is the inverse
propagator of $\psi''$, and where
\begin{equation} \label{Eq12} \Gamma_{\psi''D\bar D}(s)=
\frac{g^2_{\psi''D\bar D}}{6\pi s}\left(\frac{p^3_0(s)}{1+r^2
p^2_0(s)}+\frac{p^3_+(s)}{1+r^2p^2_+(s)}\right),
\end{equation} is the $\psi''\to D\bar D$ decay width,
where $g_{\psi''D\bar D}$ is the corresponding coupling constant.
The function $h_{\psi''}(s)$ describes the contribution of the
finite width corrections to the real part of the $\psi''$
propagator. Its explicit form is given in Appendix. Near
$s=m^2_{\psi''}$ is the function
$h_{\psi''}(s)\sim(m^2_{\psi''}-s)^2$. Values $C_{\psi''}$,
$m_{\psi''}$, $g_{\psi''D\bar D}$, and $r$ are free parameters of
the model. To normalize the form factor $F^{\psi''}_D(s)$ at
$s=m^2_{\psi''}$, we use the relation
\begin{equation}\label{Eq13} \sigma_{\psi''}(e^+e^-\to
D\bar D;\,s=m^2_{\psi''})=\frac{12\pi}{m^2_{\psi''}}\frac{
\Gamma_{\psi''e^+e^-}}{\Gamma_{\psi''D\bar D}}\,,\end{equation}
where $\Gamma_{\psi''D\bar D}\equiv\Gamma_{\psi''D\bar
D}(m^2_{\psi''})$. Then, taking into account Eqs. (\ref{Eq3}),
(\ref{Eq11}), and (\ref{Eq13}), we have (up to a sign),
\begin{equation}\label{Eq14}
C_{\psi''}=\sqrt{\frac{9m^5_{\psi''}\,\Gamma_{\psi''e^+e^-}\Gamma_{
\psi''D\bar D}}{\ 2\alpha^2\left(p^3_0(m^2_{\psi''})+p^3_+ (m^2_{
\psi''})\right)}}\ .\end{equation} Putting, by definition, $\Gamma_{
\psi''e^+e^-}=4\pi\alpha^2g^2_{\psi''\gamma}/(3m_{\psi''}^3)$, where
the constant $g_{\psi''\gamma}$ describes the $\psi''$ coupling with
the virtual $\gamma $ quantum, we can write $C_{\psi ''}$ in the
form:
\begin{equation}\label{Eq15} C_{\psi''}=g_{\psi''\gamma}\,g^{eff}_{
\psi''D\bar D}\,.
\end{equation}
The effective coupling constant of the $\psi''$ resonance with
$D\bar D$ \, $ g^{eff}_{\psi''D\bar D}$ is related to the constant
$g_{\psi''D\bar D}$ from Eq. (\ref {Eq12}) by the relation
\begin{equation}\label{Eq16}g^{eff}_{\psi''D\bar D}=\sqrt{6\pi
m^2_{\psi''}\Gamma_{\psi''D\bar D}/[p^3_0 (m^2_{\psi''})+ p^3_+
(m^2_{\psi''})]}\,.\end{equation}

From Eqs. (\ref{Eq11}) and (\ref{A1})--(\ref{A4}) it follows that,
owing to the finite width corrections in $\widetilde{D}_{\psi''}
(s)$, the form factor $F^{\psi''}_D(s)$ has good analytical
properties. In particular, it has no any singularities associated
with the poles of the functions $1/[1+r^2p^2_{0,+}(s)]$. In
addition, in $F^{\psi''}_D(s)$ there are absent spurious bound
states in the region $0<s<4m^2_{D^+}$ for $r\geq0.87$ GeV$^{-1}$
(0.174 fm) [i.e., $\widetilde{D}_{\psi''}(s)$ does not vanish
anywhere in this region].

\begin{figure} 
\begin{center}\includegraphics[width=8cm]{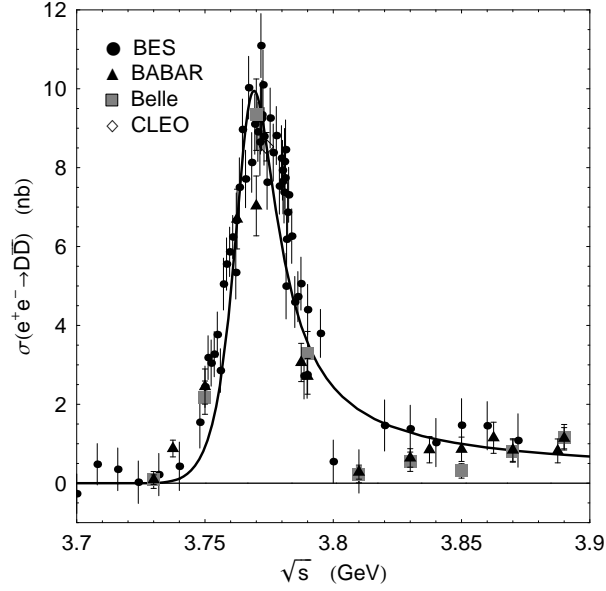}
\caption{\label{Fig1} The variant of the solitary $\psi''$ resonance
model. The curve is the fit using Eqs. (\ref{Eq3}) and
(\ref{Eq11})--(\ref{Eq15}) with the data from BES \cite{Ab3a,Ab4},
CLEO \cite{Be2}, $BABAR$ \cite{Au1,Au2}, and Belle \cite{Pa}
Collaborations for $\sigma(e^+ e^-$\,$\to$\,$D\bar D)$. There are 87
points in the fit. For more details on the data see the text and
also Ref. \cite{AS12}. }\end{center}\end{figure}

The fit to the data \cite{Ab3a,Ab4,Be2,Au1,Au2,Pa} with the use of
the solitary $\psi''$ resonance model at a fixed value of $r=0.87$
GeV$^{-1}$ is shown in Fig. \ref{Fig1}. It corresponds to $m_{
\psi''}=3.772$ GeV, $g_{\psi'' D\bar D}=14.4$ [i.e., $\Gamma_{
\psi''D\bar D}(m^2_{\psi''}) \approx27.6$ MeV], and $g_{\psi''
\gamma}=0.245$ GeV$^2$ (i.e., $\Gamma_{\psi''e^+e^-}\approx0.25$
keV). Although the obtained values of the $\psi''$ parameters are
close to those given by the PDG \cite{PDG20}, the fit in itself is
unsatisfactory. The corresponding $\chi^2=459$ for 84 degrees of
freedom. As $r$ increases, the fit becomes even less satisfactory.

With regard to the selected data \cite{Ab3a,Ab4,Be2,Au1,Au2,Pa} (see
Fig. \ref{Fig1}), we note the following. These data are the most
detailed and accurate available data on the so-called Born cross
section (i.e., on the cross section undistorted by initial state
radiation). Note that the BES Collaboration \cite{Ab3a,Ab4}
measured, in the region up to the $D\bar D^*$ threshold
($\approx$\,3.872\,GeV), the quantity $R(s)$\,=\,$\sigma (e^+e^-\to
hadrons)/\sigma(e^+e^- \to\mu^+\mu^-)$. The $D\bar D$ events were
not specially identified. The 62 BES points shown in Fig. 1
correspond to the cross section $(4\pi \alpha^2/3s )[R(s)-
R_{uds}]$, where $R_{uds}=2.121$ \cite{Ab4} describes the background
from the light hadron production. This cross section gives a good
estimate for $\sigma(e^+ e^-$\,$\to$\,$D\bar D)$ in the $\psi(3770)$
region, see the discussion in the Introduction and also in Ref.
\cite{AS12}. The utilized approximation is not critical for our
analysis.

\subsection{{\boldmath The $\psi(2S)$ contribution}}

Let us write the contribution of the state $\psi(2S)$ to $F^0_D(s)$
by analogy with Eq. (\ref {Eq11}) in the form
\begin{equation}\label{Eq17} F^0_D(s)=
F^{\psi(2S)}_D(s)=\frac{C_{\psi(2S)}}{\widetilde{D}_{\psi(2S)}(s)}=
\frac{C_{\psi(2S)}}{m^2_{\psi(2S)}-s-h_{\psi(2S)}(s)-i\sqrt{s}
\Gamma_{\psi(2S)D\bar D}(s)}\,,\end{equation} where
$m_{\psi(2S)}=3.6861$ GeV \cite{PDG20}. $F^{\psi(2S)}_D(s)$ is
calculated according to Eqs. (\ref{Eq12}) and
(\ref{A1})--(\ref{A4}), where the index $\psi''$ should be replaced
everywhere by $\psi(2S)$. The constant $C_{\psi(2S)}$ in Eq.
(\ref{Eq17}) can be represented by analogy with Eq. (\ref{Eq15}) in
the form
\begin{eqnarray}\label{Eq18} C_{\psi(2S)}=g_{\psi(2S)\gamma}\,g^{eff}_{
\psi(2S)D\bar D}\,. \end{eqnarray} The constant $g_{\psi(2S)\gamma}
$ describes the $\psi(2S)$ coupling with the virtual $\gamma$
quantum. From the PDG data \cite{PDG20}, $\Gamma_{\psi(2S)e^+e^-}=
2.33$ keV, and the relation $\Gamma_{\psi(2S)e^+e^-}=4\pi\alpha^2
g^2_{\psi (2S)\gamma}/(3m^3_{\psi(2S)})$; thus we get
$g_{\psi(2S)\gamma} =\pm0.723$ GeV$^2$.
As a free parameter for the $\psi(2S)$ contribution, it is
convenient to use the proportionality coefficient $z$ between the
coupling constants of the $\psi(2S)$ and $\psi''$ with $D\bar D$:
\begin{equation}\label{Eq19} g_{\psi(2S)D\bar
D}=z\,g_{\psi''D\bar D}\ \ \ \mbox{and}\ \ \ g^{eff}_{\psi(2S)D\bar
D}=z\,g^{eff}_{\psi''D\bar D}.
\end{equation} The relation between $g_{\psi''D\bar D}$
and $g^{eff}_{\psi''D\bar D}$ is definite by Eq. (\ref{Eq16}).

\subsection{{\boldmath $D$ meson form factor for the mixed $\psi''$ and $\psi(2S)$ states}}

We now take into account the mixing of $\psi''$ and $\psi(2S)$
resonances due to their common decay channels into $D^0\bar D^0$ and
$D^+D^-$. The form factor $F^0_D(s)$ corresponding to such a
$\psi''-\psi(2S)$ resonance complex can be written as
\cite{AS12,AS13}
\begin{equation}\label{Eq20}
F^0_D(s)=\frac{C_{\psi''}\Delta_{\psi(2S)}(s)+C_{\psi(2S)}
\Delta_{\psi''}(s)}{\widetilde{D}_{\psi''}(s)\widetilde{D}_{\psi
(2S)}(s)-\widetilde{\Pi}^2_{\psi''\psi(2S)}(s)}\,,
\end{equation}where
\begin{eqnarray}\label{Eq21}\Delta_{\psi(2S)}(s)=\widetilde{D}_{
\psi(2S)}(s)+z\,\widetilde{\Pi}_{\psi''\psi(2S)}(s)\,,\\
\label{Eq22} \Delta_{\psi''}(s)=\widetilde{D}_{\psi''}(s)
+z^{-1}\,\widetilde{\Pi}_{\psi''\psi(2S)}(s)\,,\ \ \end{eqnarray}
and $\widetilde{\Pi}_{\psi''\psi(2S)}(s)$ is the nondiagonal
polarization operator describing the transition $\psi''
$\,$\to$\,$D\bar D$\,$\to$\,$\psi(2S)$. The polarization operator
$\widetilde{\Pi}_{\psi''\psi(2S)}(s)$ is related to the diagonal
polarization operator $\Pi_{\psi''}(s)$ (see Appendix) by the
relation
\begin{eqnarray}\label{Eq23}\widetilde{\Pi}_{\psi''\psi(2S)}(s)=z\,\Pi_{
\psi''}(s)+a+s\,b,\end{eqnarray} where $a$ and $b$ are unknown
constants. In order to use the parameters introduced above for the
description of solitary $\psi''$ and $\psi(2S)$ resonances (fixed
$m_{\psi(2S)}$ and $g_{\psi(2S)\gamma}$ and free $m_{\psi''}$,
$g_{\psi''\gamma}$, $g_{\psi''D\bar D}$, and $g_{\psi(2S)D\bar D}$
or $z$) and preserve the meaning of individual characteristics for
resonances dressed by mixing, we fix the constants $a$ and $b$ by
the conditions
\begin{eqnarray}\label{Eq24}\mbox{Re}\,\widetilde{\Pi}_{
\psi''\psi(2S)}(m^2_{\psi(2S)})=0\,,\\ \label{Eq25}
\mbox{Re}\,\widetilde{\Pi}_{\psi''\psi(2S)}(m^2_{\psi''}) =0\,.\ \
\end{eqnarray}
Note that Eq. (\ref{Eq25}) keeps the normalization condition
(\ref{Eq13}) for the form factor $F^0_D(s)$ given by formula
(\ref{Eq20}). Using Eqs. (\ref{Eq24}) and (\ref{Eq25}), we find
\begin{eqnarray}\label{Eq26}
\widetilde{\Pi}_{\psi''\psi(2S)}(s)=z\,\left[\Pi_{\psi''}(s)
-\mbox{Re}\,\Pi_{\psi''}(m^2_{\psi''})+\frac{s-m^2_{\psi''}}{m^2_{
\psi''}-m^2_{\psi(2S)}}\,\mbox{Re}\left(\Pi_{\psi''} (m^2_{
\psi(2S)})-\Pi_{\psi''}(m^2_{\psi''})\right)\right].\end{eqnarray}

Note that the phase of the form factor $F^0_D(s)$, due to the strong
resonant interaction of $D$ mesons, is determined by the phase of
the denominator in Eq. (\ref{Eq20}). The numerator in this formula
is the first-degree polynomial in $s$ with real coefficients. It is
interesting that in the case under consideration we are faced,
perhaps for the first time, with the possibility of the existence of
zero in the form factor in the elastic region. As seen from Fig.
\ref{Fig1}, the data do not contradict the presence of zero in
$F^0_D(s)$ at $\sqrt{s}\approx3.81$ GeV \cite{FN3}.

\begin{figure} [!ht] 
\begin{center}\includegraphics[width=8cm]{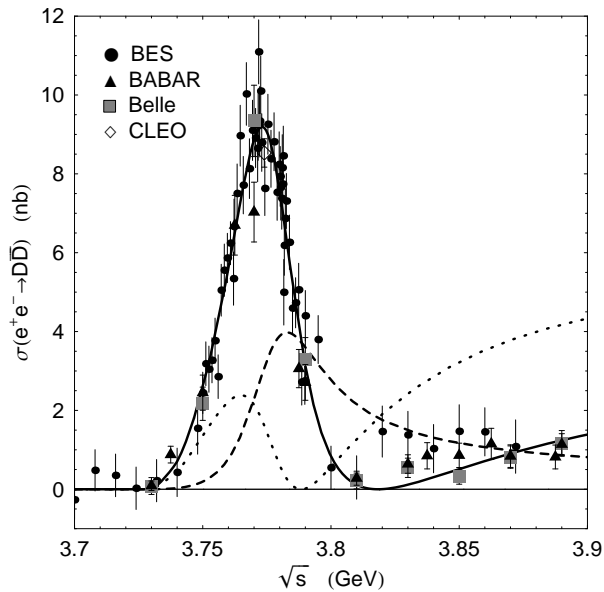}
\caption{\label{Fig2} The model of the mixed $\psi''$ and $\psi(2S)$
resonances. The solid curve is the fit using Eqs. (\ref{Eq3}) and
(\ref{Eq20})--(\ref{Eq26}) to the data from BES \cite{Ab3a,Ab4},
CLEO \cite{Be2}, $BABAR$ \cite{Au1,Au2}, and Belle \cite{Pa}
Collaborations. The dashed and dotted curves show the contributions
to the cross section from the $\psi''$ and $\psi(2S)$ production
amplitudes proportional to the coupling constants $C_{\psi''}$ and
$C_{\psi(2S)}$, respectively; see Eq. (\ref{Eq20}).}\end{center}
\end{figure}
\begin{figure} [!ht] 
\begin{center}\includegraphics[width=13.35cm]{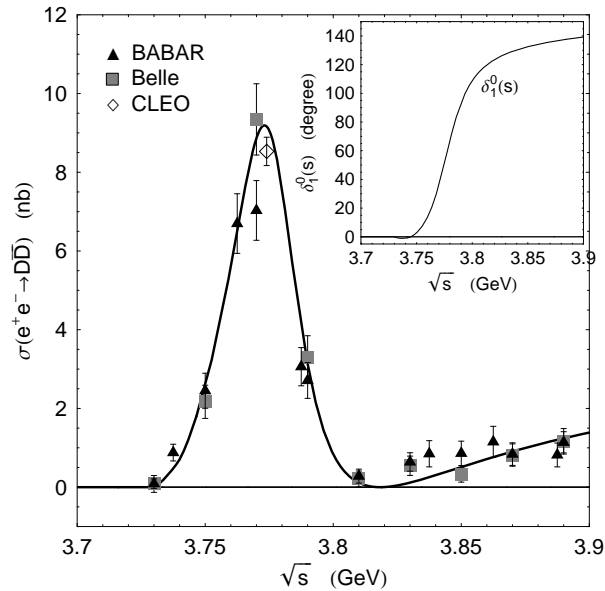}
\caption{\label{Fig3} The model of the mixed $\psi''$ and $\psi(2S)$
resonances. The curve is the same as the solid curve in Fig. 2, but
in comparison only with the data from CLEO \cite{Be2}, $BABAR$
\cite{Au1,Au2}, and Belle \cite{Pa} Collaborations. The inset shows
the phase $\delta^0_1(s)$ of the form factor $F^0_D(s)$ and $D\bar
D$ elastic scattering amplitude $T^0_1(s)$ for our fit.}\end{center}
\end{figure}

Figures \ref{Fig2} and \ref{Fig3} show the fitting of the data
\cite{Ab3a,Ab4,Be2,Au1,Au2,Pa} in the model of the mixed $\psi''$
and $\psi(2S)$ resonances. The curves in these figures correspond to
the following values of the fitted parameters: $m_{\psi''}=3.7884$
GeV, $g_{\psi''D\bar D}=60.54$,\, $g_{\psi''\gamma}=-0.2148$
GeV$^2$, and $z=1.0225$. Using these values we get $g^{eff}_{
\psi''D\bar D}=14.72$,\, $\Gamma_{\psi''D\bar D}=51.88$ MeV, and
$\Gamma_{\psi''e^+e^-}=0.189$ keV. The errors in the values of free
parameters do not exceed 5\%. For this fit, $\chi^2=127.6$, which is
approximately 3.6 times less than $\chi^2$ for the fit with the
solitary $\psi''$ resonance shown in Fig. \ref {Fig1}. Note that the
above value of the width $\Gamma_{\psi''D\bar D}$ is approximately
two times larger than the average value of the total decay width of
$\psi''$ presented by the PDG. The found mass of $\psi''$ is also 15
MeV larger than the average PDG value. However, there is not any
contradiction here. The fact is that the parameters of the $\psi''$
resonance mixed with the background cannot be directly compared with
the PDG values obtained without taking mixing into account.
Incidentally, confirmation of the existence of zero in the form
factor $F^0_D(s)$ (see Figs. \ref{Fig2} and \ref{Fig3}) would be the
best evidence that the observed peak in the region of 3.773 GeV is
the result of the interaction between the resonance and background
contributions.

The above fit in the model of the mixed $\psi''$ and $\psi (2S)$
resonances has been obtained at the fixed value of the parameter
$r=12.5$ GeV$^{- 1}$ ($\approx2.5$ fm). Let us discuss this
parameter in more detail. Its role in the description of the $\psi
''$ resonance with formulas (\ref{Eq1}) and (\ref{Eq2}) was
discussed in the second section of Ref. \cite{AS12}. Here, a few
words about $r$ were said in the two paragraphs after Eq.
(\ref{Eq16}). In Table I, we have collected the conclusions about
the parameter $r$ obtained in the processing of the data on the
$\psi (3770)$ resonance to illustrate the real situation. The
parameter $r$ is practically always taken into account when
processing resonance data, but, as a rule, it remains not
well-defined and is often simply fixed by hand. Perhaps, its main
role is to suppress the increase of the $P$-wave decay width
$\psi''\to D\bar D$ as $\sqrt{s }$ increases, see Eq. (\ref{Eq12}).
The suppression occurs faster at higher $r$. But if the fit improves
as $r$ increases, then it simultaneously becomes less sensitive to
$g^2_{\psi''D\bar D}$ and $r^2$ separately, and increasingly depends
on the ratio $g^2_{\psi''D\bar D}/r^2$ [see Eq. (\ref{Eq12})]. In
such a case the parameter $r$ remains formally unbounded from above
\cite{AS12}. With the sequential increase of $r$, one can estimate
its value, after which the $\chi^2$ of the fitting actually remains
constant. Our fit corresponds to such an approximate value of $r$.
If $r$ is decreased, then $\chi^2$ will increase, but not
catastrophically. For example, $\chi^2$ turns out to be $\approx
130.4$ at $r=5$ GeV$^{-1}$ ($\approx1$ fm). In this case
$\Gamma_{\psi''e^+e^-}\approx0.14$ keV, $\Gamma_{\psi''D \bar
D}\approx92.2$ MeV, and $m_{\psi''}\approx3.796$ GeV. Increasing the
data accuracy would make it possible to determine the value of $r$
more accurately and with it the values of other model parameters,
too.

\begin{table} 
\caption{Information about the parameter $r$ from the $\psi(3770)\to
D\bar D$ decay descriptions (1 fm $\approx5$ GeV$^{-1}$).}
\vspace{0.05cm} 
\begin{tabular}{|l|l|} \hline
  \ Data processing\,   & $\quad\quad$\  Presented conclusions \\ \hline
  \ Rapidis \cite{Ra}   & \ Acceptable fits for all values of \\
                        & \ $r>1$ fm; illustration at $r=3$ fm \\
  \ Peruzzi \cite{Pe}   & \ $r$ was varied from 0 to $\infty$ \\
  \ Schindler \cite{Sch}& \ $r$ was taken to be 2.5 fm \\
  \ Ablikim \cite{Ab1}  & \ $r$ was taken to be 0.5 fm \\
  \ Ablikim \cite{Ab2}  & \ $r$ was left free in the fit \\
  \ Ablikim \cite{Ab3}  & \ $r$ was taken to be 1 fm \\
  \ Ablikim \cite{Ab4}  & \ $r$ was a free parameter in the fit \\
  \ Ablikim \cite{Ab5}  & \ $r$ was fixed at 3 fm \\
  \ Ablikim \cite{Ab6}  & \ $r$ was of the order of a few fm \\
  \ Ablikim \cite{Ab7}  & \ $r$ was fixed at 1.5 fm \\
  \ Dobbs   \cite{Do}   & \ $r$ was taken to be 2.4 fm \\
  \ Anashin \cite{An2}  & \ $r$ was fixed at 1 fm \\
  \ Achasov \cite{AS12} & \ Analysis of Eqs. (\ref{Eq1}) and (\ref{Eq2}) for \\
                        & \ $0<r<4,...$ fm \\
  \hline\end{tabular}\end{table}

One can also express the hope that the model will become more
flexible and will improve the data description, if at the next step
of the research we take into account the couplings of the $\psi''$
and $\psi(2S)$ resonances with the closed $D\bar D^*$ and $D^*\bar
D^*$ decay channels in the region $\sqrt{s}$ up to 3.872 GeV and the
inelastic effects caused by them for $\sqrt{s}>3.872$ GeV. Of
course, further accurate measurements of the $e^+e^-\to D\bar D$
cross sections will be decisive for the selection of
phenomenological models and understanding the $\psi(3770)$ resonance
as a charm factory.

\section{{\boldmath Comparison with theoretical estimates and conclusions}}

Theoretical estimates of the electronic width of the $\psi''$
resonance, that is mainly considered the $1^3D_1$ charmonium state,
show that it is very sensitive to the relativistic corrections, QCD
corrections, and mixing of $S-D$ $c\bar c$ configurations due to
tensor forces and transitions via $D\bar D$ coupled-channels
\cite{Ei75,Ei76,Ei04,Ei80,Ei06,Ei08,No78,Ro01,
Ro05,Ja77,HTO84,So20,Bh18,Kh18}. The literature cited here presents
a rather wide range of theoretical values for $\Gamma_{\psi''
e^+e^-}$. For example, in the nonrelativistic limit, $\Gamma_{\psi
(3770)e^+e^-}$ turns out to be $\approx0.070$ keV due to the
$2^3S_1-1^3D_1$ mixing in the coupled-channel scheme \cite{Ei80}.
$\Gamma_{\psi''e^+e^-}$ can increase to $\approx0.160$ keV
\cite{Ei80}, if one takes into account the relativistic corrections
(i.e., the inequality to zero of the second derivative of the radial
wave function at the origin \cite{No78}), and further to $\approx0.
230$ keV with the connection of the the $S-D$ mixing due to tensor
forces \cite{Ei80}. The relativistic corrections (without mixing)
give for $\Gamma_{\psi''e^+e^-}$, for example, $\approx0.120$ keV
\cite{No78} or $\approx0.060$ keV \cite{Ro01}. The recent
theoretical schemes did not give more definite predictions for the
width: $\Gamma_{\psi''e^+e^-} \approx0.091$ keV \cite{So20},
$\approx0.270$ keV \cite{Bh18}, $\approx0.113$ keV \cite{Kh18}.

The spread of theoretical estimates for the width, $\Gamma_{\psi
''e^+e^-}$, quite agrees with the spread of its values found in
various experiments \cite{PDG20} and also in accompanying
phenomenological analyses \cite{Ya,LQY,ZZ,AS12,AS13,CZ,Li,CL14a,
CL14b,DMW,ST,CG} (see discussions in previous sections). Of course,
the primary guide is the value of $\Gamma_{\psi''e^+e^-}=(0.262
\pm0.018)$ keV given by the PDG \cite{PDG20}. However, as noted
above, the phenomenological formulas used to obtain this value were
rather simplified (or even poorly grounded). If the errors of the
data on $\sigma(e^+e^-\to D\bar D)$ are reduced by approximately two
times compared to the existing ones [see Figs. (\ref{Fig2}) and
(\ref{Fig3})], then it will be possible to abandon such formulas.
When processing new, more accurate data on the cross section
$\sigma(e^+e^- \to D^0\bar D^0+D^+D^-)$, it will make sense to take
into account the Coulomb interaction in the final state between
$D^+$ and $D^-$ mesons, which amplifies the charged channel by about
8.8\% at the peak of the $\psi''$ resonance \cite{La77}.


Now we summarize: 1) The model of the $D$ meson form factor
$F^0_D(s)$ with good unitary and analytic properties is constructed
to describe the cross section of the reaction $e^+e^-\to D\bar D$
near the threshold, 2) The model involves the complex of the mixed
$\psi''$ and $\psi(2S)$ resonances and satisfactorily describes the
data in the $\sqrt{s}$ region up to 3.9 GeV, 3) A feature of the
model is the presence of zero in $F^0_D(s)$ at $\sqrt{s}\approx
3.818$ GeV, 4) The survey of the experimental, phenomenological, and
theoretical results for $\Gamma_{\psi''e^+e^-}$ is also presented to
illustrate the variety of approaches to determining this quantity,
and 5) The rather small value of $\Gamma_{\psi''e^+e^-}\approx0.19$
keV, obtained by us, and the corresponding value of the ratio
$\Gamma_{\psi''e^+e^-}/\Gamma_{\psi(2S)e^+e^-}\approx0.081$ indicate
in favor of the $D$-wave $c\bar c$ nature of the $\psi''$ state.

Improving the data on the shape of the $\psi(3770)$ resonance in
the $D\bar D$ decay channels seems to be an extremely important and
quite feasible physical problem.\\

\begin{center} {\bf ACKNOWLEDGMENTS} \end{center}

The work was carried out within the framework of the state contract
of the Sobolev Institute of Mathematics, Project No.
0314-2019-0021.\\

\begin{center} {\bf APPENDIX:\, THE FUNCTION {\boldmath
$h_{\psi''}(s)$}} \end{center}


\setcounter{equation}{0}
\renewcommand{\theequation}{A\arabic{equation}}

The twice subtracted dispersion integral corresponding to the
one-loop $P$-wave Feynman diagram has the form:
\begin{eqnarray}\label{A1} f_{0,+}(s)=\frac{s^2}{\pi}
\int\limits^\infty_{4m^2_{D^{0,+}}}\frac{p^3_{0,+}(s')\,ds'}{\sqrt{s'}
\,s'^2(s'-s-i\varepsilon)}& \nonumber\\ =\frac{s-3m^2_{D^{0,+}}
}{3\pi}-\frac{s\rho_{0,+}(s)^3}{8\pi}\ln\frac{\rho_{0,+}(s)+1
}{\rho_{0,+}(s)-1}\,,&\quad &\mbox{for\ } s<0\,, \nonumber\\
=\frac{s-3m^2_{D^{0,+}}}{3\pi}+\frac{s| \rho_{0,+} (s)|^3}{8\pi
}\left(\pi-2\arctan |\rho_{0,+}(s)|\right),& \quad & \mbox{for\ }
0<s<4m^2_{D^{0,+}}\,,\nonumber\\ =\frac{s-3m^2_{D^{0,+}}}{
3\pi}+\frac{s\rho^3_{0,+}(s)}{ 8\pi}\left(i\pi-\ln\frac{
1+\rho_{0,+}(s)}{1-\rho_{0,+} (s)}\right),&\quad &\mbox{for\ }
s>4m^2_{D^{0,+}}\,,
\end{eqnarray} where
$\rho_{0,+}(s)=2p_{0,+}(s)/\sqrt{s}=\sqrt{1-4m^2_{D^{0,+}}/s}\,$.
The polarization operators of the $\psi''$ resonance
$\Pi^0_{\psi''}(s)$ and $\Pi^+_{\psi''}(s)$ corresponding to the
contributions of the $D^0\bar D^0$ and $D^+D^-$ intermediate states
are expressed in terms of the functions $f_{0}(s)$ and $f_{+}(s)$ as
follows:
\begin{eqnarray}\label{A2}
\Pi^{0,+}_{\psi''}(s)=\frac{g^2_{\psi''D\bar D}}{6\pi}
\,\frac{s^2}{\pi}\int\limits^\infty_{4m^2_{D^{0,+}}}\frac{p^3_{0,+}
(s')\,ds'}{\sqrt{s'}\,(1+r^2p^2_{0,+}(s'))\,s'^2(s'-s-i\varepsilon)
}\nonumber\\ =\frac{g^2_{\psi''D\bar D}}{6\pi}\frac{1}{1+r^2p^2_{
0,+}(s)}\left[ f_{0,+}(s)-\left( \frac{s}{s_{0,+}}\right)^2f_{0,+}
(s_{0,+})\right],\quad
\end{eqnarray} where $s_{0,+}=4(m^2_{D^{0,+}}-1/r^2)$.
The knowledge of the $\psi''$ mass squared, $m^2_{\psi''}$, and the
$\psi''$ width at $s=m^2_{\psi''}$, $\Gamma_{\psi''D\bar D}$, allows
us to represent the function $h_{\psi''}(s)$, entering in Eq.
(\ref{Eq11}), in the form \cite{GS,RP,BM}:
\begin{eqnarray}\label{A3}
h_{\psi''}(s)=\mbox{Re}\Pi_{\psi''}(s)-\mbox{Re}\Pi_{\psi''}(m^2_{\psi''})-
(s-m^2_{\psi''})\mbox{Re}\Pi'_{\psi''}(m^2_{\psi''}), \qquad
\end{eqnarray} where $\Pi_{\psi''}(s)=\Pi^0_{\psi''}
(s)+\Pi^+_{\psi''}(s)$ is the full polarization operator of
$\psi''$,
\begin{eqnarray}\label{A4}\mbox{Im}\Pi_{\psi''}(s)=\sqrt{s}\Gamma_{\psi''D\bar D}
(s).\end{eqnarray} See Eqs. (\ref{Eq11}) and (\ref{Eq12}).



\begin{thebibliography}{99}
\bibitem{FN1} Important examples of the form factors described by solitary resonances were constructed in the works \cite{GS,RP,BM}.
\bibitem{FN3} The possible position of zero in $F^0_D(s)$ is consistent with the prediction of the coupled-channel
              model \cite{Ei80}.
\end{thebibliography}
\end{document}